\documentclass[10pt,fleqn]{revtex4}
\usepackage[latin9]{inputenc}
\usepackage{xcolor}
\usepackage{pdfcolmk}
\usepackage{amsmath}
\usepackage{amssymb}
\usepackage{graphicx}
\usepackage{esint}
\PassOptionsToPackage{normalem}{ulem}
\usepackage{ulem}
\usepackage[unicode=true]
 {hyperref}

\makeatletter

\DeclareRobustCommand{\greektext}{%
  \fontencoding{LGR}\selectfont\def\encodingdefault{LGR}}
\DeclareRobustCommand{\textgreek}[1]{\leavevmode{\greektext #1}}
\DeclareFontEncoding{LGR}{}{}
\DeclareTextSymbol{\~}{LGR}{126}
\providecommand{\tabularnewline}{\\}
\providecolor{lyxadded}{rgb}{0,0,1}
\providecolor{lyxdeleted}{rgb}{1,0,0}

\@ifundefined{textcolor}{}
{%
 \definecolor{BLACK}{gray}{0}
 \definecolor{WHITE}{gray}{1}
 \definecolor{RED}{rgb}{1,0,0}
 \definecolor{GREEN}{rgb}{0,1,0}
 \definecolor{BLUE}{rgb}{0,0,1}
 \definecolor{CYAN}{cmyk}{1,0,0,0}
 \definecolor{MAGENTA}{cmyk}{0,1,0,0}
 \definecolor{YELLOW}{cmyk}{0,0,1,0}
}

\@ifundefined{date}{}{\date{}}
\newcommand{\Htwop}{\hbox{\rm H}_2^+}
\newcommand{\Dtwop}{\hbox{\rm D}_2^+}

\newcommand{\HDp}{\hbox{\rm HD}^+}

\makeatother

\begin{document}

\title{The static and dynamic polarisability, and the Stark and black-body
radiation frequency shifts of the molecular hydrogen ions H$_{2}^{+}$,
HD$^{+}$, and D$_{2}^{+}$ }

\author{S. Schiller}

\affiliation{\textit{Institut für Experimentalphysik, Heinrich-Heine-Universität
Düsseldorf, 40225 Düsseldorf, Germany}}

\author{D. Bakalov }

\affiliation{\textit{Institute for Nuclear Research and Nuclear Energy, Tsarigradsko
chaussée 72, Sofia 1784, Bulgaria }}

\author{A.~K. Bekbaev }

\affiliation{Al-Farabi Kazakh National University,050012, Almaty, Kazakhstan }

\author{V.~I. Korobov}

\affiliation{\textit{Joint Institute for Nuclear Research, 141980, Dubna, Russia}}
\begin{abstract}
We calculate the DC Stark effect for three molecular hydrogen ions
in the non-relativistic approximation. The effect is calculated both
in dependence on the rovibrational state and in dependence on the
hyperfine state. We discuss special cases and approximations. We also
calculate the AC polarisabilities for several rovibrational levels,
and therefrom evaluate accurately the black-body radiation shift,
including the effects of excited electronic states. The results enable
the detailed evaluation of certain systematic shifts of the transitions
frequencies for the purpose of ultra-high-precision optical, microwave
or radio-frequency spectroscopy in ion traps.
\end{abstract}
\maketitle

\section{Introduction}

The molecular hydrogen ions represent a family of simple quantum systems
that are amenable both to high-precision ab-initio calculations \cite{Gremaud98,Korobov Hilico Karr 2014}
and to high-precision spectroscopy. Therefore, they are of great interest
for the determination of fundamental constants \cite{Schiller 2007},
for tests of the time- and gravitational-potential-independence of
fundamental constants \cite{SS1,SS2}, and for tests of QED \cite{Korobov Hilico Karr 2014}.
On the experimental side, after early pioneering work on uncooled
trapped ions and ions beams \cite{Jefferts,Wing,Carrington}, the
sympathetic cooling of trapped molecular hydrogen ions \cite{Schiller03,Blythe 2005}
has opened up the window for high-precision radio-frequency, rotational,
and rovibrational spectroscopy. Precision infrared laser spectroscopy
of two rovibrational transitions has been achieved \cite{Schiller 2007,Bressel 2012},
and the fundamental rotational transition has also been observed \cite{Shen}.

Because of the advances in experimental accuracy, and in order to
open perspectives for future work directions, it has become important
to evaluate the systematic effects on the transition frequencies.
It is an advantage of the molecular hydrogen ion family that the sensitivities
to external fields can be calculated ab-initio. The systematic effects
treated so far include the Zeeman shift \cite{Zee1,Zee2,Karr Korobov Hilico Zeeman effect in H2+},
the electric quadrupole shift \cite{Bakalov and Schiller 2013}, and
the black-body radiation (BBR) shift \cite{kool}. The electric polarisability
of the rovibrational levels of the molecular hydrogen ions has been
of interest for a long time. It was computed with high accuracy for
a subset of levels by several authors, in particular \cite{Bishop and Lam 1988,Bhatia-1,Hilico 2001,moss,Yan,Karr Kilic Hilico 2005}.
These calculations used adiabatic or non-adiabatic wave functions.
Ref.~\cite{Hilico 2001} reviews the experimental and theoretical
values for the ground state of $\Htwop$ and $\Dtwop$. A particularly
accurate calculation of the polarisability of $\Htwop$ in its ground
state was performed by one of the present authors, by including the
relativistic corrections \cite{KorPRA00}. The dependence of the polarisability
on the hyperfine state has only recently been obtained \cite{hfi},
for the case of $\HDp$, and it was shown that the dependence is very
significant. These results have permitted a first analysis of the
potential for ultra-high accuracy spectroscopy of $\HDp$ and its
suitability as an optical clock \cite{Bakalov and Schiller 2013,Schiller Bakalov Korobov PRL 2014}.

In the present paper, extensive calculations of the polarisability
are presented. Its dependence on the hyperfine state is derived in
a more elegant way and discussed in depth, both for $\HDp$ and $\Htwop$,
since it is of great relevance for experiments.

While the BBR shift is tiny, it will eventually become of relevance
for experiments requiring the highest levels of accuracy, such as
the mentioned test of time-independence of fundamental constants.
Therefore, this shift is also computed in detail. Results for $\Htwop$
are presented for the first time. In addition, the case of $\HDp$
is treated extensively, in view of the current experimental interest
in this molecule.

This paper is structured as follows: In Sec.~2 we briefly review
the calculation approach for the polarisability of the molecular hydrogen
ions, neglecting spin effects. We define the effective Hamiltonian
and present the tables of polarisabilities. In Sec.~3 we introduce
the hyperfine structure and discuss the computation of the DC Stark
shift in dependence of the spin state. We also give a number of useful
approximations. Sec.~4 presents detailed results for a large number
of hyperfine states potentially relevant for high-precision spectroscopy.
In Sec.~5 we discuss the energy level shifts induced by the oscillating
(AC) electric field of the black-body radiation, which we accurately
evaluate by taking into account the precise frequency dependence of
the polarisability.

\section{Evaluation of the polarisability}

\subsection{Non-relativistic polarisability. Spin-independent spatial considerations}

For the purposes of evaluating the systematic effects in spectroscopy,
it is at present sufficient to use the non-relativistic approximation
to the polarisability. Therefore, we start from the non-relativistic
Schrödinger equation:

\begin{equation}
(H_{0}-E)\Psi_{0}=0,\qquad H_{0}=-\frac{1}{2M_{1}}\mbox{\ensuremath{\nabla}}_{1}^{2}-\frac{1}{2M_{2}}\mbox{\ensuremath{\nabla}}_{2}^{2}-\frac{1}{2m_{e}}\mbox{\ensuremath{\nabla}}^{2}+\frac{1}{R}-\frac{1}{r_{1}}-\frac{1}{r_{2}},\label{eq:NR}
\end{equation}
where $M_{1}$ and $M_{2}$ are the masses of the nuclei (proton or
deuteron), $R$ is the internuclear distance, $r_{1}$ and $r_{2}$
are the distances from nuclei 1 and 2 to the electron, respectively.
The state $\Psi_{0}=|v\, L\rangle$ is the unperturbed state characterized
by the vibrational and rotational quantum numbers $v,\, L$, and $E_{0}$
is its energy.

The interaction with an external electric field ${\bf E}$ in the
dipole interaction form is expressed by

\begin{equation}
V_{p}=-\mathbf{E}\cdot{\bf d},\qquad{\bf d}=e\left[Z(\mathbf{R}_{1}+\mathbf{R}_{2})-\mathbf{r}\right]\,,
\end{equation}
where $\mathbf{d}$ is the electric dipole moment of the three-particle-system,
${\bf R}_{1,2}$ and $\mathbf{r}$ are the position vectors of the
nuclei and of the electron with respect to the center of mass.

Since the static or quasi-static electric fields present in an ion
trap, and also the electric field of the radiation from continuous-wave
lasers and from the black-body environmental radiation are typically
weak, it is sufficient to apply second-order perturbation theory for
the calculation of the polarisability. The energy shifts that result
are typically at the level of 1~Hz, orders of magnitude smaller than
the rotational or hyperfine splittings. For effects of higher-order
in the external electric field, see Ref.~\cite{Bhatia-1}.

The change of energy due to the polarisability of a molecular ion
is expressed by

\begin{eqnarray}
E_{p}^{(2)} & = & \langle\Psi_{0}|V_{p}(E_{0}-H_{0})^{-1}V_{p}|\Psi_{0}\rangle\nonumber \\
 & = & E^{i}E^{j}\langle\Psi_{0}|d^{i}(E_{0}-H_{0})^{-1}d^{j}|\Psi_{0}\rangle\\
 & = & -\frac{1}{2}\alpha_{d}^{ij}E^{i}E^{j}\,,\nonumber
\end{eqnarray}
where $\alpha_{d}^{ij}$, the polarisability tensor of rank 2, has
been introduced,

\begin{equation}
\alpha_{d}^{ij}=-2\langle\Psi_{0}|d^{i}(E_{0}-H_{0})^{-1}d^{j}|\Psi_{0}\rangle.
\end{equation}
The static dipole polarisability tensor is then reduced to scalar,
$\alpha_{s}$, and tensor, $\alpha_{t}$, terms, which may be expressed
in terms of three contributions corresponding to the possible values
of of the rotational angular momentum quantum number of the intermediate
state: $L'=L\pm1$, or $L'=L$.

\begin{eqnarray}
a_{+} & = & \;\;\frac{2}{2L+1}\sum_{p}\frac{\langle v\, L\|\mathbf{d}\|p\,(L+1)\rangle\langle p\,(L+1)\|\mathbf{d}\|v\, L\rangle}{E_{0}-E_{p}}\,,\nonumber \\
a_{\text{0}} & = & -\frac{2}{2L+1}\sum_{p}\frac{\langle v\, L\|\mathbf{d}\|p\, L\rangle\langle p\, L\|\mathbf{d}\|v\, L\rangle}{E_{0}-E_{p}}\,,\nonumber \\
a_{-} & = & \;\;\frac{2}{2L+1}\sum_{p}\frac{\langle v\, L\|\mathbf{d}\|p\,(L-1)\rangle\langle p\,(L-1)\|\mathbf{d}\|v\, L\rangle}{E_{0}-E_{p}}\,,\label{eq:a+ a0 and a-}
\end{eqnarray}
where $L$ is the rotational quantum number of the state under consideration,
$E_{p}$ is the non-relativistic energy of the intermediate state
$|p\, L'\rangle$.

The polarisability tensor may be expressed as

\begin{equation}
\alpha_{d}^{ij}=\delta_{ij}\alpha_{s}+\alpha_{t}\left\langle \Psi_{0}|L_{i}L_{j}+L_{j}L_{i}-\frac{2}{3}\delta_{ij}\mathbf{L}^{2}|\Psi_{0}\right\rangle \,,\label{eq:alpha:op}
\end{equation}
where $L_{i}$ are the Cartesian components of the rotational angular
momentum operator, $\mathbf{L}^{2}=L_{x}^{2}+L_{y}^{2}+L_{z}^{2}$,
and

\begin{eqnarray*}
\alpha_{s} & = & \frac{1}{3}(a_{+}+a_{0}+a_{-}),\\
\alpha_{t} & = & -\frac{a_{+}}{2(L+1)(2L+3)}+\frac{a_{0}}{2L(L+1)}-\frac{a_{-}}{2L(2L-1)}\,.
\end{eqnarray*}

We may also define longitudinal ($\alpha_{\parallel}$) and transverse
($\alpha_{\bot}$) polarisabilities

\begin{align}
\alpha_{\parallel} & =\alpha_{s}+\alpha_{t}\left\langle \Psi_{0}|2\, L_{z}^{2}-\frac{2}{3}\mathbf{L}^{2}|\Psi_{0}\right\rangle \\
\alpha_{\bot} & =\frac{1}{2}(\alpha_{d}^{xx}+\alpha_{d}^{yy})=\alpha_{s}+\alpha_{t}\left\langle \Psi_{0}|L_{x}^{2}+L_{y}^{2}-\frac{2}{3}\mathbf{L}^{2}|\Psi_{0}\right\rangle =\alpha_{s}-\frac{1}{2}\alpha_{t}\left\langle \Psi_{0}|2\, L_{z}^{2}-\frac{2}{3}\mathbf{L}^{2}|\Psi_{0}\right\rangle .
\end{align}

 The definition of $\alpha_{\perp}$ as given above is reasonable,
since axial symmetry requires that the matrix elements of $L_{x}^{2}$
and of $L_{y}^{2}$ are equal. Thus, the polarisabilities $\alpha_{\parallel}$
and $\alpha_{\perp}$ actually involve the expectation value of only
a single operator, which has an alternative representation as the
0-component of the rank-2 tensor $\{{\bf L}\otimes{\bf L}\}_{2}$,

\begin{equation}
2\, L_{z}^{2}-\frac{2}{3}\mathbf{L}^{2}=\sqrt{\frac{8}{3}}\{{\bf L}\otimes{\bf L}\}_{20}\,.
\end{equation}
In Sec.~\ref{sec:Perturbation-theory-for}, we will evaluate the
polarisabilities of the hyperfine states of a given ro-vibrational
level. The approximation we will use consists in introducing the polarisability
operator, which holds in a manifold of given $L$,

\begin{eqnarray*}
\hat{\alpha}_{d}^{ij}(v,\, L) & = & \alpha_{s}(v,\, L)+\alpha_{t}(v,\, L)\left[L_{i}L_{j}+L_{j}L_{i}-\frac{2}{3}\delta_{ij}\mathbf{L}^{2}\right]\,,\\
\hat{\alpha}_{\parallel}(v,\, L) & = & \alpha_{s}(v,\, L)+\alpha_{t}(v,\, L)\sqrt{\frac{8}{3}}\{{\bf L}\otimes{\bf L}\}_{20}\,,\\
\hat{\alpha}_{\perp}(v,\, L) & = & \alpha_{s}(v,\, L)-\frac{1}{2}\alpha_{t}(v,\, L)\sqrt{\frac{8}{3}}\{{\bf L}\otimes{\bf L}\}_{20}\,.
\end{eqnarray*}
Here, we have included the explicit dependence of the coefficients
$\alpha_{s}$, $\alpha_{t}$ on the vibrational and rotational quantum
numbers $v,\, L$. In the following, we will explicitly consider the
polarisability anisotropy operator,

\begin{equation}
\hat{\alpha}_{\parallel}-\hat{\alpha}_{\perp}=\alpha_{t}(v,\, L)\,\frac{3}{2}\,\sqrt{\frac{8}{3}}\{{\bf L}\otimes{\bf L}\}_{20}=3\alpha_{t}(v,\, L)\left(L_{z}^{2}-\frac{1}{3}\mathbf{L}^{2}\right)\,.\label{eq:definition of polarization anisotropy}
\end{equation}

\subsection{Numerical results}

Wave functions of the rovibrational states in the molecular hydrogen
ions are obtained by using the variational approach expounded in Ref.~\cite{KorPRA00}.
Briefly, the wave function for a state with a total orbital angular
momentum $L$ and of a total spatial parity $\pi=(-1)^{L}$ is expanded
as follows:

\begin{eqnarray}
\Psi_{LM}^{\pi}(\mathbf{R},\mathbf{r}_{1}) & = & \sum_{l_{1}+l_{2}=L}\mathcal{Y}_{LM}^{l_{1}l_{2}}(\hat{\mathbf{R}},\hat{\mathbf{r}}_{1})G_{l_{1}l_{2}}^{L\pi}(R,r_{1},r_{2}),\nonumber \\
G_{l_{1}l_{2}}^{L\pi}(R,r_{1},r_{2}) & = & \sum_{n=1}^{N}\Big\{ C_{n}\,\mbox{Re}\bigl[e^{-\alpha_{n}R-\beta_{n}r_{1}-\gamma_{n}r_{2}}\bigr]+\nonumber \\
 &  & \,\,\,\,\,\,\, D_{n}\,\mbox{Im}\bigl[e^{-\alpha_{n}R-\beta_{n}r_{1}-\gamma_{n}r_{2}}\bigr]\Big\}\,,
\end{eqnarray}
where the complex exponents $\alpha$, $\beta$, $\gamma$, are generated
in a pseudo-random way. The use of complex exponents instead of real
ones allows reproducing the oscillatory behavior of the vibrational
part of the wave function and improves the convergence rate. In numerical
calculations we utilize basis sets as large as $N$ = 7~000 functions
, in order to provide the required accuracy for the static polarisability
of about 8 significant digits.

We note that a variational principle holds for the numerical value
for $\alpha_{s}$ (but not for $\alpha_{t}$): the larger the value,
the closer it is to the exact (non-relativistic) value, provided that
the initial wave function is accurate enough.

The results of numerical calculations of the polarisabilities for
a wide range of ro-vibrational states are presented in Tables 1,~2,~3.
These polarisabilities do not include relativistic corrections. These
have so far been computed only for the ground rovibrational level
$(v=0,\, L=0)$ of $\Htwop$ \cite{KorPRA00}. Therefore, the relative
inaccuracy of the values of the table as compared to the exact values
is of order $\alpha^{-2}\simeq1\times10^{-4}$. This is sufficiently
small for current and near-future purposes.

\begin{table}
\begin{centering}
\caption{\label{table: polarisability of HD+}Polarisabilities of the $\mbox{HD}^{+}$
molecular ion, in atomic units.}

\par\end{centering}

\centering{}$\begin{tabular}{|r|@{\hspace{3mm}}c|r@{\hspace{3mm}}c|r@{\hspace{3mm}}c|r@{\hspace{3mm}}r|r@{\hspace{3mm}}c|r@{\hspace{3mm}}c|ccccccc}
\hline\hline   &  \ensuremath{L=0} &  \ensuremath{L=1} &   &  \ensuremath{L=2} &   &  \ensuremath{L=3} &   &  \ensuremath{L=4} &   &  \ensuremath{L=5} &  \\
\ensuremath{v} &  \ensuremath{\alpha_{s}} &  \ensuremath{\alpha_{s}~} &  \ensuremath{\alpha_{t}} &  \ensuremath{\alpha_{s}~} &  \ensuremath{\alpha_{t}} &  \ensuremath{\alpha_{s}~} &  \ensuremath{\alpha_{t}~} &  \ensuremath{\alpha_{s}~} &  \ensuremath{\alpha_{t}} &  \ensuremath{\alpha_{s}~} &  \ensuremath{\alpha_{t}}\\
\hline 0  &  395.30633  &  3.99015  &  175.48275  &  4.00956  &  13.82797  &  4.03878  &  3.19075  &  4.07794  &  1.10141  &  4.12721  &  0.47319 \\
1  &  462.65271  &  4.70314  &  205.20067  &  4.72694  &  16.14340  &  4.76278  &  3.71557  &  4.81084  &  1.27799  &  4.87136  &  0.54642 \\
2  &  540.68636  &  5.56925  &  239.58035  &  5.59871  &  18.81611  &  5.64313  &  4.31921  &  5.70273  &  1.48001  &  5.77786  &  0.62955 \\
3  &  631.40288  &  6.63284  &  279.47585  &  6.66965  &  21.91000  &  6.72541  &  5.01516  &  6.80017  &  1.71152  &  6.89451  &  0.72396 \\
4  &  737.31802  &  7.95478  &  325.95893  &  8.00132  &  25.50477  &  8.07195  &  5.82011  &  8.16691  &  1.97742  &  8.28690  &  0.83127 \\
5  &  861.64968  &  9.61839  &  380.39514  &  9.67856  &  29.70139  &  9.76943  &  6.75494  &  9.89175  &  2.28374  &  10.04654  &  0.95337 \\
6  &  1008.5802  &  11.74323  &  444.54814  &  11.82178  &  34.62944  &  11.94052  &  7.84610  &  12.10056  &  2.63789  &  12.30342  &  1.09241 \\
7  &  1183.6432  &  14.50032  &  520.73882  &  14.60466  &  40.45801  &  14.76254  &  9.12757  &  14.97563  &  3.04910  &  15.24624  &  1.25088 \\
8  &  1394.3075  &  18.14238  &  612.07821  &  18.28368  &  47.41173  &  18.49776  &  10.64364  &  18.78717  &  3.52889  &  19.15548  &  1.43147 \\
9  &  1650.8846  &  23.05215  &  722.82833  &  23.24788  &  55.79504  &  23.54473  &  12.45301  &  23.94684  &  4.09171  &  24.45984  &  1.63690 \\
10  &  1967.9875  &  29.82774  &  858.97404  &  30.10584  &  66.03006  &  30.52844  &  14.63477  &  31.10210  &  4.75562  &  31.83608  &  1.86935
\\\hline\hline \end{tabular}$
\end{table}

\begin{table}
\small
\begin{centering}
\caption{\label{table: polarisability of H2+}Polarisabilities of the $\mbox{H}_{2}^{+}$
molecular ion, in atomic units.}

\par\end{centering}

\centering{}$\begin{tabular}{|r|@{\hspace{1mm}}c|c@{\hspace{2mm}}c|c@{\hspace{2mm}}c|c@{\hspace{2mm}}c|c@{\hspace{2mm}}c|c@{\hspace{2mm}}c|ccccccccccc}
\hline\hline   &  \ensuremath{L=0} &  \ensuremath{L=1} &   &  \ensuremath{L=2} &   &  \ensuremath{L=3} &   &  \ensuremath{L=4} &   &  \ensuremath{L=5} &  \\
\ensuremath{v} &  \ensuremath{\alpha_{s}} &
\ensuremath{\alpha_{s}} &  \ensuremath{\alpha_{t}} &
\ensuremath{\alpha_{s}} &  \ensuremath{\alpha_{t}} &
\ensuremath{\alpha_{s}} &  \ensuremath{\alpha_{t}} &
\ensuremath{\alpha_{s}} &  \ensuremath{\alpha_{t}} &
\ensuremath{\alpha_{s}} &  \ensuremath{\alpha_{t}}\\
\hline 0  &  3.1687258  &  3.1783035  &  -0.8033729  &  3.1975081  &  -0.1931423  &  3.2264392  &  -0.0914467  &  3.2652493  &  -0.0544769  &  3.3141473  &  -0.0367142 \\
1  &  3.8975634  &  3.9101018  &  -1.1442051  &  3.9352574  &  -0.2751013  &  3.9731892  &  -0.1302653  &  4.0241411  &  -0.0776138  &  4.0884471  &  -0.0523179 \\
2  &  4.8215004  &  4.8380889  &  -1.6000689  &  4.8713900  &  -0.3847653  &  4.9216560  &  -0.1822373  &  4.9892726  &  -0.1086157  &  5.0747693  &  -0.0732474 \\
3  &  6.0093275  &  6.0315483  &  -2.2129563  &  6.0761862  &  -0.5322759  &  6.1436389  &  -0.2521973  &  6.2345166  &  -0.1503892  &  6.3496578  &  -0.1014845 \\
4  &  7.5604532  &  7.5906530  &  -3.0434869  &  7.6513642  &  -0.7322875  &  7.7432180  &  -0.3471422  &  7.8671844  &  -0.2071498  &  8.0246002  &  -0.1399110 \\
5  &  9.6217735  &  9.6635170  &  -4.1811566  &  9.7475033  &  -1.0064626  &  9.8747452  &  -0.4774336  &  10.046804  &  -0.2851555  &  10.265837  &  -0.1928182 \\
6  &  12.416000  &  12.474853  &  -5.7615823  &  12.593371  &  -1.3876723  &  12.773211  &  -0.6588274  &  13.016932  &  -0.3939491  &  13.328069  &  -0.2667729 \\
7  &  16.290999  &  16.375936  &  -7.9965515  &  16.547168  &  -1.9273337  &  16.807463  &  -0.9160304  &  17.161118  &  -0.5485440  &  17.614095  &  -0.3721509 \\
8  &  21.809473  &  21.935532  &  -11.228720  &  22.189990  &  -2.7087984  &  22.577626  &  -1.2892120  &  23.105870  &  -0.7734466  &  23.785138  &  -0.5259729 \\
9  &  29.920328  &  30.113886  &  -16.036300  &  30.505195  &  -3.8730473  &  31.102847  &  -1.8465559  &  31.920266  &  -1.1104555  &  32.976407  &  -0.7574477 \\
10  &  42.306330  &  42.616316  &  -23.445884  &  43.244200  &
-5.6711124  &  44.206257  &  -2.7100058 &  45.528094 & -1.6347702
&  47.246181  & -1.1195247
\\\hline\hline \end{tabular}$
\end{table}

\begin{table}
\begin{centering}
\caption{\label{table: polarisability of D2+} Polarisabilities of the $\mbox{D}_{2}^{+}$
molecular ion, in atomic units.}

\par\end{centering}

\centering{}$\begin{tabular}{|r|@{\hspace{1mm}}c|c@{\hspace{1mm}}c|c@{\hspace{2mm}}c|c@{\hspace{2mm}}c|c@{\hspace{2mm}}c|c@{\hspace{2mm}}c|cccccccccc}
\hline\hline   &  \ensuremath{L=0} &  \ensuremath{L=1} &   &  \ensuremath{L=2} &   &  \ensuremath{L=3} &   &  \ensuremath{L=4} &   &  \ensuremath{L=5} &  \\
\ensuremath{v} &  \ensuremath{\alpha_{s}} &
\ensuremath{\alpha_{s}} &  \ensuremath{\alpha_{t}} &
\ensuremath{\alpha_{s}} &  \ensuremath{\alpha_{t}} &
\ensuremath{\alpha_{s}} &  \ensuremath{\alpha_{t}} &
\ensuremath{\alpha_{s}} &  \ensuremath{\alpha_{t}} &
\ensuremath{\alpha_{s}} &  \ensuremath{\alpha_{t}}\\
\hline 0  &  3.0719887  &  3.0765904  &  -0.7579521  &  3.0858052  &  -0.1813435  &  3.0996560  &  -0.0852443  &  3.1181777  &  -0.0503016  &  3.1414173  &  -0.0335048 \\
1  &  3.5530258  &  3.5585822  &  -0.9782731  &  3.5697111  &  -0.2340592  &  3.5864444  &  -0.1100266  &  3.6088309  &  -0.0649271  &  3.6369364  &  -0.0432481 \\
2  &  4.1195817  &  4.1263238  &  -1.2485988  &  4.1398301  &  -0.2987476  &  4.1601453  &  -0.1404432  &  4.1873367  &  -0.0828824  &  4.2214959  &  -0.0552137 \\
3  &  4.7912827  &  4.7995087  &  -1.5808716  &  4.8159913  &  -0.3782716  &  4.8407920  &  -0.1778439  &  4.8740043  &  -0.1049671  &  4.9157545  &  -0.0699367 \\
4  &  5.5933149  &  5.6034134  &  -1.9904009  &  5.6236531  &  -0.4763025  &  5.6541185  &  -0.2239603  &  5.6949390  &  -0.1322078  &  5.7462891  &  -0.0881048 \\
5  &  6.5583187  &  6.5708021  &  -2.4970077  &  6.5958274  &  -0.5975951  &  6.6335113  &  -0.2810366  &  6.6840342  &  -0.1659357  &  6.7476365  &  -0.1106108 \\
6  &  7.7290547  &  7.7446049  &  -3.1266348  &  7.7757864  &  -0.7483752  &  7.8227607  &  -0.3520126  &  7.8857801  &  -0.2078964  &  7.9651778  &  -0.1386263 \\
7  &  9.1622096  &  9.1817469  &  -3.9136471  &  9.2209342  &  -0.9368936  &  9.2799934  &  -0.4407873  &  9.3592871  &  -0.2604068  &  9.4592730  &  -0.1737083 \\
8  &  10.933925  &  10.958708  &  -4.9041723  &  11.008431  &  -1.1742306  &  11.083398  &  -0.5526002  &  11.184144  &  -0.3265836  &  11.311297  &  -0.2179539 \\
9  &  13.147977  &  13.179752  &  -6.1610454  &  13.243527  &  -1.4754879  &  13.339760  &  -0.6946003  &  13.469145  &  -0.4106838  &  13.632624  &  -0.2742310 \\
10  &  15.948121  &  15.989359  &  -7.7712809  &  16.072159  &
-1.8615919  &  16.197178  &  -0.8766990 &  16.365416 & -0.5186172
&  16.578236  & -0.3465280
\\\hline\hline \end{tabular}$
\end{table}

\subsection{Scaling with rotational angular momentum}

For large $L$, we find for $\HDp$,

\begin{equation}
\alpha_{t}(v,\, L)\varpropto\frac{1}{L(L+1)(2L-1)(2L+3)}\,.\label{eq:approximate polarisability of HD+}
\end{equation}
This follows from an argument described below after Eq.~(\ref{eq:anisotropic polarisability any pure state}).

We have found heuristically, that for $\Htwop$ and $\Dtwop$,

\begin{equation}
\alpha_{t}(v,\, L)\varpropto\frac{1}{(2L-1)(2L+3)}\,.
\end{equation}

\subsection{Comparison with previous work}

\subsubsection{Contribution from the ground electronic state}

An approximation to the polarisability can be obtained using the well-known
sum-over-intermediate-states expression, where the sum is truncated
to a subset of levels. For $\HDp$, such a calculation has been performed
using transition dipole moments computed in the Born-Oppenheimer approximation
\cite{hfi}, including in the sum only levels of the ground electronic
state (the inaccuracy of the used dipole moments is mentioned further
below). At first sight, it may appear that the polarisability of a
level $(v,\, L)$ in $\HDp$ is dominated by the contribution from
the rotational levels adjacent in energy to the particular state,
namely $(v,\, L\pm1)$. This is evidently true for $L=0$ levels.
However, for $L\ne0$, there is partial cancellation of the two contributions
from $L'=L\pm1$. Even then, the $\alpha_{t}$ values indeed arise
essentially from the rovibrational transitions. However, the $\alpha_{s}(L>0)$
values are actually dominated by the contribution from the excited
electronic states. The comparison of the accurate results given in
the tables above with the truncated-sum results allows putting in
evidence the contribution from the excited electronic states. The
comparison is shown in Table~\ref{tab:Comparison-of-the polarisabilities},
showing that for low-lying rovibrational levels ($v,\, L<5)$, the
difference is of order several atomic units for $\alpha_{s}$ and
less than 2.5 atomic units for $\alpha_{t}$. The increase of the
difference with $v$ is due to the fact that the contributions from
excited electronic states become more important since the level $v$
is getting closer in energy to them.

\begin{table}
\centering{}$\begin{array}{|c|c|cc|cc|cc|cc|}
\hline \text{\ensuremath{v}} & L=0 & L=1 &  & L=2 &  & L=3 &  & L=4 & \\
\hline \text{} & \delta\alpha_{s} & \delta\alpha_{s} & \delta\alpha_{t} & \delta\alpha_{s} & \delta\alpha_{t} & \delta\alpha_{s} & \delta\alpha_{t} & \delta\alpha_{s} & \delta\alpha_{t}\\
\hline 0 & 3.3 & 3.1 & -0.72 & 3.1 & -0.18 & 3.2 & -0.08 & 3.2 & -0.06\\
1 & 3.9 & 3.7 & -0.98 & 3.8 & -0.24 & 3.8 & -0.12 & 3.8 & -0.08\\
2 & 4.7 & 4.5 & -1.34 & 4.5 & -0.34 & 4.6 & -0.16 & 4.6 & -0.10\\
3 & 5.6 & 5.4 & -1.80 & 5.5 & -0.44 & 5.5 & -0.22 & 5.6 & -0.12\\
4 & 6.8 & 6.6 & -2.42 & 6.6 & -0.60 & 6.7 & -0.28 & 6.8 & -0.16
\\\hline \end{array}$\caption{\label{tab:Comparison-of-the polarisabilities} Difference $\delta\alpha$
between the accurate polarisabilities of $\HDp$ (this work) and those
computed by a summation over all intermediate rovibrational states
in the ground electronic state, in atomic units. The latter are calculated
from the results of Ref.~\cite{hfi} (which are there given in terms
of $\alpha_{vLFSJJ_{z}}^{(l)},$ $\alpha_{vLFSJJ_{z}}^{(t)}$) as
$\alpha_{s}(v,\, L)=(\alpha_{vLFSJJ_{z}}^{(l)}+2\alpha_{vLFSJJ_{z}}^{(t)})/3$
(where any hyperfine state $(F,\, S,\, J,\, J_{z})$ can be chosen)
and $\alpha{}_{t}(v,\, L)=(\alpha_{vLn_{s}}^{(l)}-\alpha_{vLn_{s}}^{(t)})/L(2L-1)$,
where $n_{s}$ denotes the stretched state, defined in Sec.~\ref{sub:The-stretched-states}. }
\end{table}

For the homonuclear $\Htwop$ and $\Dtwop$ the polarisability arises
only from the excited electronic states, since there is no electric-dipole
coupling between levels of the ground electronic state. As a consequence,
the polarisabilities $\alpha_{s}(L=0)$ and $\alpha_{t}(v,\, L)$
are much smaller than in the case of the heteronuclear ions, as has
been noted in previous studies cited above.

\subsubsection{General calculations}

We can compare our results with some previous studies.

Early on, Bishop and Lam \cite{Bishop and Lam 1988} studied the states
$v=0,\, L=0-10$ of $\Htwop$. The largest number of levels was considered
by Moss and Valenzano, who covered the three ion species also treated
here, with $L=0,\,1$, and all $v$ \cite{moss}. Our results agree
with theirs, to within two units of the last digit reported by them,
except for the level $(v=8,\, L=1)$, where the largest discrepancy
occurs, 0.007 at.~u.

The agreement with the $L=0$ values for the three ion species determined
by Hilico et al. \cite{Hilico 2001}, and Karr et al. \cite{Karr Kilic Hilico 2005}
is better than $4\times10^{-8}$ in relative terms.

Pilon and Baye recently computed the polarisabilities of $\Htwop$
for a number of levels \cite{Pilon and Baye}. The values for $v=0,\, L=0,\,1,\,2,\,3,\,4,\,5$
agree to better than $2\times10^{-8}$ in relative terms. For $L=1,\, v=0,\,1,\,2,\,3$
and for $L=2,\, v=0,\,1,\,2,\,3$ the values agree with the present
values to better than $3\times10^{-7}$ in relative terms.

\section{Perturbation theory for the hyperfine states\label{sec:Perturbation-theory-for}}

\subsection{Energy shifts}

The hyperfine interactions split each rovibrational level into a number
of hyperfine sub-levels. We denote the corresponding kets as $|m\rangle=|v\, L\, n\, J_{z}\rangle$,
where $n$ is a label for the particular hyperfine state in a rovibrational
level $(v,\, L)$ (note that this notation includes both pure and
non-pure spin states). $n$ is written as $(F,\, S,\, J)$ for $\HDp$
and $(I,\, S,\, J)$ for $\Htwop$, see Sec.~\ref{sub:Hyperfine-structure}
below. When the Stark shifts of the quantum levels are small compared
to other shifts, we can apply first-order perturbation theory. The
Stark energy shift of a state $|m\rangle$ can be expressed in different
ways (for simplicity, in the following we omit the caret on the polarization
operators) \cite{Arosa presentation}:

\begin{align}
\Delta E(m) & =-\frac{1}{2}\left[\langle m|\alpha_{\parallel}|m\rangle E_{z}^{2}+\langle m|\alpha_{\perp}|m\rangle(E_{x}^{2}+E_{y}^{2})\right]\,,\nonumber \\
 & =-\frac{1}{2}\left[\alpha_{s}(E_{x}^{2}+E_{y}^{2}+E_{z}^{2})+\alpha_{t}(E_{z}^{2}-\frac{1}{2}(E_{x}^{2}+E_{y}^{2}))\langle m|\sqrt{\frac{8}{3}}\{{\bf L}\otimes{\bf L}\}_{20}|m\rangle\right]\,,\nonumber \\
 & =-\frac{1}{2}\alpha_{s}(E_{x}^{2}+E_{y}^{2}+E_{z}^{2})-\alpha_{t}(E_{z}^{2}-\frac{1}{2}(E_{x}^{2}+E_{y}^{2}))\langle m|L_{z}^{2}-\frac{1}{3}{\bf L}^{2}|m\rangle\,,\nonumber \\
 & =-\frac{1}{2}{\bf E}^{2}\left(\alpha_{s}+\alpha_{t}(3\cos^{2}\theta-1)\langle m|L_{z}^{2}-\frac{1}{3}{\bf L}^{2}|m\rangle\right)\,,\nonumber \\
 & =-\frac{1}{2}{\bf E}^{2}\left(\alpha_{s}(v,\, L)+(3\cos^{2}\theta-1)\frac{1}{3}\langle v\, L\, n\, J_{z}|\alpha_{\parallel}-\alpha_{\perp}|v\, L\, n\, J_{z}\rangle\right)\,,\label{eq:Stark shift general}
\end{align}
where $\theta$ is the angle between the quantization axis and the
direction of the electric field ${\bf E}$.

In Refs.~\cite{hfi,Bakalov and Schiller 2013} the levels shifts
were described in terms of longitudinal polarisability, $\alpha^{(l)}$,
and transverse polarisability $\alpha^{(t)}$. They are related to
the expectation values of the operators introduced here by $\langle m|\alpha_{\parallel}|m\rangle=\alpha^{(l)}$
and $\langle m|\alpha_{\perp}|m\rangle=\alpha^{(t)}$.

\subsection{Hyperfine structure\label{sub:Hyperfine-structure}}

We limit ourselves in the following to the ion species $\Htwop$ and
$\HDp$, which are most relevant for experimental work at present.

In case of the molecular ion $\Htwop$ we have identical nuclei and
nuclear permutation symmetry. This makes some spin configurations
forbidden and splits the consideration of hyperfine states into two
cases (see \cite{KorPRA06R}): for even $L$, the total nuclear spin
$I$ is zero and only two hyperfine sub-levels are possible; for states
with odd $L$, the total nuclear spin is one and the ro-vibrational
level is split into 5 or 6 hyperfine sub-levels, depending on the
value of $L$.

The most suitable coupling scheme of angular momentum operators is

\begin{equation}
\mathbf{S}=\mathbf{I}+\mathbf{s}_{e},\qquad\mathbf{J}=\mathbf{S}+\mathbf{L\,},\label{eq:couplinf:H2}
\end{equation}
where $\mathbf{I}$ is the total nuclear spin operator, and ${\bf s}_{e}$
is the electron spin operator. The basis states which correspond to
this coupling are

\begin{equation}
|ISLJJ_{z}\rangle=\sum_{I_{z},\,\xi,\, S_{z}}C_{S\, S_{z},L\, L_{z}}^{J\, J_{z}}C_{I\, I_{z},s_{e}\xi}^{S\, S_{z}}\Bigl(|I\, I_{z}\rangle\cdot|s_{e}\xi\rangle\cdot|L\, L_{z}\rangle\Big)\,,\label{eq:purestates H2plus}
\end{equation}
and will be called pure states \cite{Karr Bielsa et al 2008}.

The effective HFS Hamiltonian is expressed as \cite{KorPRA06R}

\begin{equation}
\begin{array}{@{}l}
{\displaystyle H_{{\rm eff}}=b_{F}(\mathbf{I}\cdot\mathbf{s}_{e})+c_{e}(\mathbf{L}\cdot\mathbf{s}_{e})+c_{I}(\mathbf{L}\cdot\mathbf{I})}\\[2mm]
{\displaystyle \hspace{10mm}+\frac{d_{1}}{(2L-1)(2L+3)}\biggl\{\frac{2}{3}\mathbf{L}^{2}(\mathbf{I}\cdot\mathbf{s}_{e})-\Bigl[(\mathbf{L}\cdot\mathbf{I})(\mathbf{L}\cdot\mathbf{s}_{e})+(\mathbf{L}\cdot\mathbf{s}_{e})(\mathbf{L}\cdot\mathbf{I})\Bigr]\biggr\}}\\[3mm]
{\displaystyle \hspace{10mm}+\frac{d_{2}}{(2L-1)(2L+3)}\left[\frac{1}{3}\mathbf{L}^{2}\mathbf{I}^{2}-\frac{1}{2}(\mathbf{L}\cdot\mathbf{I})-(\mathbf{L}\cdot\mathbf{I})^{2}\right].}
\end{array}\label{eq:effH_H2}
\end{equation}
For the case of even $L$, the pure states are the true HFS eigenstates,
since the $2\!\times\!2$ effective HFS Hamiltonian matrix is diagonal.
Even for odd-$L$ states, the pure states are good approximations
to the true HFS states \cite{Karr Bielsa et al 2008}, since the coefficients
of admixture of other states to a given true HFS state are small,
e.g. for $L=1$ do not exceed 0.04, and for $L=3$ do not exceed 0.06.
This means that even in this case, a good approximation for expectation
values such as Eq.~(\ref{eq:Stark shift general}) may be obtained
using the pure states.

For the hydrogen molecular ion $\HDp$ the coupling scheme of the
particle angular momentum operators is \cite{prl06}

\begin{equation}
\mathbf{F}=\mathbf{I}_{p}+\mathbf{s}_{e},\quad\mathbf{S}=\mathbf{F}+\mathbf{I}_{d},\quad\mathbf{J}=\mathbf{S}+\mathbf{L}\,.\label{eq:coupling:HD}
\end{equation}
${\bf I}_{p},\,{\bf I}_{d}$ are the proton and deuteron spin operators,
respectively. The effective Hamiltonian is given in Ref.~\cite{prl06}.
The pure states are determined in a similar way as in Eq.~(\ref{eq:purestates H2plus}).
In zero magnetic field, the pure states represent a good approximation
to some of the true HFS states, and may be used to calculate approximate
values of the polarisabilities. Details are given in Sec.~\ref{sec:Numerical-results-for the hyperfine state dependence}
below. Hyperfine states are labeled by $n=(F\, S\, J)$.

\subsection{Analytical results}

In this subsection we discuss some useful results that allow to understand
several dependencies. In particular we discuss the polarisabilities
of the pure spin states, for two reasons. First, a significant part
of hyperfine states may be well approximated by pure spin states;
second, since all hyperfine states can be expressed as weighted sums
of pure spin states, their polarisabilities can conveniently be computed
from the pure state polarisabilities.

\subsubsection{Zero magnetic field}

When the magnetic field is zero, the total angular momentum squared
${\bf J}^{2}$ commutes with the hyperfine Hamiltonian and $J$ is
a good quantum number. Therefore we can apply the Wigner-Eckart theorem,
and separate the $J_{z}$- dependence of the expectation value:

\begin{eqnarray}
\langle vLnJ_{z}|\{{\bf L}\otimes{\bf L}\}_{20}|vLnJ_{z}\rangle & = & C_{20,JJ_{z}}^{JJ_{z}}\langle vLn||\{{\bf L}\otimes{\bf L}\}_{2}||vLn\rangle/\sqrt{2J+1}\nonumber \\
 & = & -\frac{J(J+1)-3J_{z}^{2}}{\sqrt{J(J+1)(2J-1)(2J+1)(2J+3)}}\langle vLn||\{{\bf L}\otimes{\bf L}\}_{2}||vLn\rangle
\end{eqnarray}
We therefore obtain the $J_{z}$ - dependence of the polarisability
anisotropy as follows:

\begin{equation}
\langle vLnJ_{z}|\alpha_{\parallel}-\alpha_{\perp}|vLnJ_{z}\rangle=\left(1-\frac{3\, J_{z}^{2}}{J(J+1)}\right)\langle vLnJ_{z}=0|\alpha_{\parallel}-\alpha_{\perp}|vLnJ_{z}=0\rangle\,.\label{eq:Jz-dependence}
\end{equation}
Note that this result holds both for pure and non-pure spin states.
It follows that for $J=0$ states, the polarisability anisotropy is
zero. For $\HDp$, $J=0$ states can only occur for $L<3$, since
the minimum $J$ value permitted by angular momentum algebra is $L-2$.
For $\Htwop$, there are no such states, since $J$ is a half-integer
number.

\subsubsection{Pure states}

For pure angular momentum states, the matrix elements of the polarisability
anisotropy can be evaluated explicitly. Considering only the coupling
scheme $\mathbf{J}=\mathbf{S}+\mathbf{L}$, we have (note that this
is independent of $I$ or $F$)
\begin{equation}
\begin{array}{@{}l}
{\displaystyle \left\langle SLJJ_{z}\left|2\left(L_{z}^{2}-\frac{1}{3}\mathbf{L}^{2}\right)\right|SLJJ_{z}\right\rangle =\sqrt{\frac{8}{3}}\left\langle SLJJ_{z}|\{\mathbf{L}\otimes\mathbf{L}\}_{20}|SLJJ_{z}\right\rangle \,\,,}\\[4mm]
\hspace{10mm}{\displaystyle =\sqrt{\frac{8}{3}}\,\frac{C_{JJ_{z},20}^{JJ_{z}}}{\sqrt{2J+1}}\,\left\langle SLJ\|\{\mathbf{L}\otimes\mathbf{L}\}_{2}\|SLJ\right\rangle }\,\,,\\[4mm]
\hspace{10mm}{\displaystyle =-\frac{(J(J+1)-3J_{z}^{2})\,[3D(D-1)-4J(J+1)L(L+1)]}{3J(J+1)(2J-1)(2J+3)}}\,\,,
\end{array}\label{eq:matrix element for pure states}
\end{equation}

where
\begin{equation}
D=J(J+1)+L(L+1)-S(S+1)\,.\label{eq:D}
\end{equation}
This result is obtained using the following relations \cite{Messiah,Varshalovich 1988}

\[
\begin{array}{@{}l}
{\displaystyle \bigl\langle SLJ\bigl\|\{\mathbf{L}\otimes\mathbf{L}\}_{2}\bigr\| SLJ\bigr\rangle=(2J+1)(-1)^{S+L+J+2}\left\{ \begin{matrix}L & \! L\! & 2\\
J & \! J\! & S
\end{matrix}\right\} \left\langle L\|\{\mathbf{L}\otimes\mathbf{L}\}_{2}\|L\right\rangle \,,}\\[3mm]
{\displaystyle \left\langle L\|\{\mathbf{L}\otimes\mathbf{L}\}_{2}\|L\right\rangle =\frac{1}{\sqrt{6}}\sqrt{L(L+1)}\sqrt{(2L-1)(2L+1)(2L+3)}}\,.
\end{array}
\]
In $\Htwop$ we consider first the states having even $L$, so $I=0$.
Then $S=1/2$. These pure states are exact HFS eigenstates, and therefore
Eq.~(\ref{eq:matrix element for pure states}) immediately gives
the exact Stark shift using Eqs.~(\ref{eq:Stark shift general},~\ref{eq:Jz-dependence}):

\begin{equation}
\langle m({\rm even}\, L)|\alpha_{\parallel}-\alpha_{\perp}|m({\rm even}\, L)\rangle_{{\rm H_{2}^{+}}}=-\frac{3}{2}\alpha_{t}(v,\, L)\,\frac{(J(J+1)-3J_{z}^{2})\,[3D(D-1)-4J(J+1)L(L+1)]}{3J(J+1)(2J-1)(2J+3)}\,.\label{eq:anisotropic polarisability for even L for H2plus}
\end{equation}
with $D=J(J+1)+L(L+1)-3/4$.

For pure states with odd $L$ (and therefore $I=1$):

\begin{equation}
\begin{array}{@{}l}
\left\langle I=1\, S\, L\, J\Bigl\|\{\mathbf{L}\otimes\mathbf{L}\}_{2}\Bigr\| I=1\, S\, L\, J\right\rangle =(2J+1)(-1)^{S+L+J}\left\{ \begin{matrix}L & \! L\! & 2\\
J & \! J\! & S
\end{matrix}\right\} \left\langle L\|\left\{ \mathbf{L}\otimes\mathbf{L}\right\} _{2}\|L\right\rangle \\[3mm]
\hspace{10mm}{\displaystyle =\sqrt{\frac{2J+1}{6}}\,\frac{3D(D-1)-4J(J+1)L(L+1)}{2\sqrt{J(J+1)(2J-1)(2J+3)}}\,,}
\end{array}\label{eq:reduced matrix element for odd L}
\end{equation}
where $D$ is given by Eq.~(\ref{eq:D}). We see that the actual
value of $I$ does not occur on the r.h.s., and that we obtain the
same result as for the $I=0$ pure states. Eq.~(\ref{eq:anisotropic polarisability for even L for H2plus})
is an approximate result also for the odd-$L$ hyperfine states of
$\Htwop$ which are not pure, provided they are approximately pure
(see below).

In the case of $\HDp$, where the pure states are denoted as $|F\, S\, L\, J\, J_{z}\rangle$,
Eq.~(\ref{eq:anisotropic polarisability for even L for H2plus})
also holds, where $L$ now can be even or odd. There is no dependence
on $F$.

Summarizing, for any pure state of $\Htwop$ and $\HDp$, and, by
consequence, also for all other molecular hydrogen ions, Eq.~(\ref{eq:anisotropic polarisability for even L for H2plus})
gives the polarisability anisotropy:

\begin{equation}
\langle{\rm pure\, state}|\alpha_{\parallel}-\alpha_{\perp}|{\rm pure\, state}\rangle_{\text{{\rm any species}}}=-\frac{3}{2}\alpha_{t}(v,\, L)\,\frac{(J(J+1)-3J_{z}^{2})\,[3D(D-1)-4J(J+1)L(L+1)]}{3J(J+1)(2J-1)(2J+3)}\,.\label{eq:anisotropic polarisability any pure state}
\end{equation}

We note that Roeggen \cite{Roeggen} has developed an approximate
theory of the polarisability of heteronuclear diatomic molecules with
spin, neglecting nuclear spin. If we combine our Eq.~(\ref{eq:anisotropic polarisability any pure state})
and the approximate dependencies Eq.~(\ref{eq:approximate polarisability of HD+})
we reproduce the result given in Eqs.~(62,~63) of Ref.~\cite{Roeggen}.

\subsubsection{The stretched states\label{sub:The-stretched-states}}

The stretched states are those exact HFS states having maximal total
angular momentum $J$ and maximal (absolute) projection $|J_{z}|$.
These are also pure states. In $\HDp$, these are the states $|v\, L\, n_{s}\rangle$,
where $n_{s}$ denotes the stretched hyperfine state: $F=1,\, S=2,\, J=L+2,$
$J_{z}=\pm(L+2)$. We find from Eq.~(\ref{eq:anisotropic polarisability for even L for H2plus})
or by analytical evaluation of the matrix elements for these two stretched
states (the evaluation is simple, if the calculations is done with
the basis functions being the eigenfunctions of the individual angular
momenta $I_{p},\, I_{d},\, S,\, L$),
\begin{equation}
\langle v\, L\, n_{s}|\alpha_{\parallel}-\alpha_{\perp}|v\, L\, n_{s}\rangle=L(2L-1)\,\alpha_{t}(v,\, L)\,.\label{eq:Stretched states result}
\end{equation}
Compare the discussion in Ref.~\cite{Bakalov and Schiller 2013}.

In $\Htwop$, the stretched states are $|v\, L\, n_{s}\rangle\equiv|v,\, L,\, I=1,\, S=3/2,\, J=L+3/2,\, J_{z}=\pm(L+3/2)\text{\ensuremath{\rangle}}$.
The same result Eq.~(\ref{eq:Stretched states result}) is obtained.

By evaluating the polarisabilities of all hyperfine states, we find
that if $L>1$, the largest value of $\langle m|\alpha_{\parallel}-\alpha_{\perp}|m\rangle$
within a rovibrational level occurs for the stretched states (see
tables below). Therefore, in the following discussion, we normalize
the polarisability anisotropy values of any hyperfine state in a particular
rovibrational level relative to that of the stretched states in that
same level.

For $\HDp$, combining the result Eq.~(\ref{eq:Stretched states result})
for the stretched states with the approximate behavior Eq.~(\ref{eq:approximate polarisability of HD+}),
we obtain $\langle v\, L\, n_{s}|\alpha_{\parallel}-\alpha_{\perp}|v\, L\, n_{s}\rangle\propto(3+5\, L+2\, L^{2})^{-1}$.
This describes a rather strong decrease in the magnitude of all anisotropic
polarisability values, not only those of the stretched-states, with
increasing $L$.

\section{Numerical results for the hyperfine-state dependence\label{sec:Numerical-results-for the hyperfine state dependence}}

The evaluation of the matrix elements of Eq.~(\ref{eq:definition of polarization anisotropy})
for all (exact) hyperfine states is straightforward, once the hyperfine
states in absence of electric field are known. The calculation can
for example proceed by considering the expansion of the hyperfine
states in pure states, and then applying Eq.~(\ref{eq:anisotropic polarisability any pure state}),
which holds for the pure states of any molecular hydrogen ion. Actually,
the matrix elements are the same (apart from prefactors such as $\alpha_{t}$)
as the matrix elements for the electric quadrupole shift evaluated
in Ref.\ \cite{Bakalov and Schiller 2013}, and an explicit formula
is given there.

We have performed the computation for the rovibrational levels up
to $v=4$ and $L=4$. Note that the polarisability anisotropy vanishes
for $L=0$ states and is therefore not reported in the tables. We
confine ourselves to the case of zero magnetic field.

The results are summarized in table \ref{tab:The-normalized-polarisability for L=00003D1 and HD+}
and table \ref{tab:The normalised polarisabilities for L 2 3 4 for HD+}
where we give the values for the hyperfine states having $J_{z}=0$.
The values for $J_{z}\ne0$ can be easily obtained using Eq.~(\ref{eq:Jz-dependence}).
Note that for a given hyperfine state and value of $L$ the dependence
on $v$ is usually weak, limited to several percent, except for a
few cases.

\begin{table}
$\begin{array}{|l|c|c|c|c|c|c|}
\hline {\rm {\rm hyp.\, state}} & {\rm Eq.\,(}\ref{eq:anisotropic polarisability any pure state}) &  &  & {\rm level} &  & \\
\text{\ensuremath{(F,\, S,\, J,\, J_{z})}} & \text{(normalized)} & \text{\ensuremath{(v,L)=(0,1)}} & \text{\ensuremath{(v,L)=(1,1)}} & \text{\ensuremath{(v,L)=(2,1)}} & \text{\ensuremath{(v,L)=(3,1)}} & \text{\ensuremath{(v,L)=(4,1)}}\\
\hline (0,1,2,0) & -1 & -0.999422 & -0.999457 & -0.999491 & -0.999525 & -0.999558\\
(0,1,1,0) & 1 & 0.998747 & 0.998825 & 0.998902 & 0.998978 & 0.999053\\
(0,1,0,0)* & 0 & 0 & 0 & 0 & 0 & 0\\
(1,0,1,0) & -2 & -1.66845 & -1.69239 & -1.71585 & -1.73892 & -1.76151\\
(1,1,1,0) & 1 & 0.590503 & 0.619326 & 0.647628 & 0.675521 & 0.702883\\
(1,1,0,0)* & 0 & 0 & 0 & 0 & 0 & 0\\
(1,1,2,0) & -1 & -0.999973 & -0.999972 & -0.999971 & -0.99997 & -0.99997\\
(1,2,1,0) & -0.2 & -0.120803 & -0.125763 & -0.130681 & -0.135574 & -0.140428\\
(1,2,3,0)* & -0.8 & -0.8 & -0.8 & -0.8 & -0.8 & -0.8\\
(1,2,2,0) & 1 & 0.999395 & 0.999429 & 0.999462 & 0.999495 & 0.999528
\\\hline \end{array}$\caption{\label{tab:The-normalized-polarisability for L=00003D1 and HD+} The
normalized anisotropic polarisabilities for the $J_{z}=0$ - hyperfine
states of $\HDp$ in $L=1$ levels. The first column shows the hyperfine
state's label $n,\, J_{z}$ ($F$ and $S$ are usually approximate,
$J$ is an exact quantum number), the second column contains the normalized
values $\langle v\, L\, F\, S\, J\, J_{z}=0|\alpha_{\parallel}-\alpha_{\perp}|v\, L\, F\, S\, J\, J_{z}=0\rangle/\langle v\, L\, n_{s}|\alpha_{\parallel}-\alpha_{\perp}|v\, L\, n_{s}\rangle$
for the pure state $|v\, L\, F\, S\, J\, J_{z}=0\rangle$ giving the
largest contribution to the exact HFS state $|v\, L\, n\, J_{z}=0\rangle$
(note that the values are independent of $F$ and of $v$). The following
columns give the actual values $\langle v\, L\, n\, J_{z}=0|\alpha_{\parallel}-\alpha_{\perp}|v\, L\, n\, J_{z}=0\rangle/\langle v\, L\, n_{s}|\alpha_{\parallel}-\alpha_{\perp}|v\, L\, n_{s}\rangle$
for each hyperfine state. The normalization is with respect to the
polarisability anisotropy of the stretched state $|v\, L\, n_{s}\rangle$
of the same rovibrational level. The three states marked with an asterisk
are pure states. }
\end{table}

\begin{table}
$\begin{array}{|l|c|c|c|c|c|c|}
\hline {\rm hyp.\, state} & {\rm Eq.}\,(\ref{eq:anisotropic polarisability any pure state}) &  &  & {\rm level}(v,L) &  & \\
\text{\ensuremath{(F,\, S,\, J,\, J_{z})}} & \text{{\rm (normalized)}} & \text{\ensuremath{(v,L)=(0,2)}} & \text{\ensuremath{(v,L)=(1,2)}} & \text{\ensuremath{(v,L)=(2,2)}} & \text{\ensuremath{(v,L)=(3,2)}} & \text{\ensuremath{(v,L)=(4,2)}}\\
\hline (0,1,3,0) & -0.8 & -0.799487 & -0.799517 & -0.799548 & -0.799578 & -0.799608\\
(0,1,2,0) & -0.5 & -0.499349 & -0.49939 & -0.49943 & -0.499469 & -0.499508\\
(0,1,1,0) & -0.7 & -0.698944 & -0.699011 & -0.699076 & -0.699141 & -0.699205\\
(1,0,2,0) & -1 & -0.89668 & -0.902108 & -0.907625 & -0.913227 & -0.918913\\
(1,1,1,0) & -0.7 & -0.598522 & -0.605382 & -0.612152 & -0.618825 & -0.625393\\
(1,1,2,0) & -0.5 & -0.572569 & -0.56875 & -0.564853 & -0.560883 & -0.556839\\
(1,1,3,0) & -0.8 & -0.798847 & -0.798904 & -0.798961 & -0.799019 & -0.799079\\
(1,2,0,0)* & 0 & 0 & 0 & 0 & 0 & 0\\
(1,2,1,0) & 0.7 & 0.597466 & 0.604392 & 0.611229 & 0.617966 & 0.624598\\
(1,2,2,0) & 0.2143 & 0.182884 & 0.184534 & 0.186194 & 0.187865 & 0.189546\\
(1,2,3,0) & -0.2 & -0.201666 & -0.201579 & -0.201491 & -0.201403 & -0.201313\\
(1,2,4,0)* & -0.7143 & -0.714286 & -0.714286 & -0.714286 & -0.714286 & -0.714286\\
\hline \text{} & \text{} & \text{\ensuremath{(v,L)}=(0,3)} & \text{\ensuremath{(v,L)}=(1,3)} & \text{\ensuremath{(v,L)}=(2,3)} & \text{\ensuremath{(v,L)}=(3,3)} & \text{\ensuremath{(v,L)}=(4,3)}\\
\hline (0,1,4,0) & -0.7143 & -0.713807 & -0.713835 & -0.713864 & -0.713892 & -0.71392\\
(0,1,3,0) & -0.6 & -0.599383 & -0.599421 & -0.599459 & -0.599496 & -0.599533\\
(0,1,2,0) & -0.6857 & -0.684872 & -0.684925 & -0.684977 & -0.685028 & -0.685079\\
(1,0,3,0) & -0.8 & -0.743487 & -0.745805 & -0.748206 & -0.750693 & -0.753272\\
(1,1,2,0) & -0.6857 & -0.641563 & -0.644157 & -0.646756 & -0.649358 & -0.651959\\
(1,1,3,0) & -0.6 & -0.638925 & -0.637509 & -0.636019 & -0.634449 & -0.632793\\
(1,1,4,0) & -0.7143 & -0.712034 & -0.712155 & -0.712277 & -0.712399 & -0.712523\\
(1,2,1,0)* & -0.48 & -0.48 & -0.48 & -0.48 & -0.48 & -0.48\\
(1,2,2,0) & -0.1714 & -0.216421 & -0.213776 & -0.211124 & -0.208471 & -0.205819\\
(1,2,3,0) & -0.2533 & -0.271538 & -0.270598 & -0.26965 & -0.268696 & -0.267735\\
(1,2,4,0) & -0.4286 & -0.431302 & -0.431152 & -0.431002 & -0.430852 & -0.4307\\
(1,2,5,0)* & -0.6667 & -0.666667 & -0.666667 & -0.666667 & -0.666667 & -0.666667\\
\hline \text{} & \text{} & \text{\ensuremath{(v,L)}=(0,4)} & \text{\ensuremath{(v,L)}=(1,4)} & \text{\ensuremath{(v,L)}=(2,4)} & \text{\ensuremath{(v,L)}=(3,4)} & \text{\ensuremath{(v,L)}=(4,4)}\\
\hline (0,1,5,0) & -0.6667 & -0.666213 & -0.66624 & -0.666267 & -0.666293 & -0.66632\\
(0,1,4,0) & -0.6071 & -0.606572 & -0.606607 & -0.606642 & -0.606676 & -0.60671\\
(0,1,3,0) & -0.6548 & -0.654035 & -0.654081 & -0.654126 & -0.65417 & -0.654214\\
(1,0,4,0) & -0.7143 & -0.677583 & -0.678822 & -0.68012 & -0.68148 & -0.682906\\
(1,1,3,0) & -0.6548 & -0.627904 & -0.62934 & -0.630793 & -0.632265 & -0.633749\\
(1,1,4,0) & -0.6071 & -0.630532 & -0.629989 & -0.62939 & -0.628733 & -0.628016\\
(1,2,2,0)* & -0.5612 & -0.561224 & -0.561224 & -0.561224 & -0.561224 & -0.561224\\
(1,1,5,0) & -0.6667 & -0.663695 & -0.663856 & -0.664019 & -0.664183 & -0.664348\\
(1,2,3,0) & -0.3929 & -0.420442 & -0.41896 & -0.417462 & -0.415946 & -0.414418\\
(1,2,4,0) & -0.4096 & -0.423439 & -0.422708 & -0.421974 & -0.421236 & -0.420494\\
(1,2,5,0) & -0.5 & -0.503426 & -0.503237 & -0.503047 & -0.502857 & -0.502666\\
(1,2,6,0)* & -0.6364 & -0.636364 & -0.636364 & -0.636364 & -0.636364 & -0.636364
\\\hline \end{array}$\caption{\label{tab:The normalised polarisabilities for L 2 3 4 for HD+}Same
as Table~\ref{tab:The-normalized-polarisability for L=00003D1 and HD+},
but for levels $L=2,\,3,\,4$. Note that the values in the second
column are rounded. }
\end{table}

By looking at the values in the Tables~\ref{tab:The-normalized-polarisability for L=00003D1 and HD+},~\ref{tab:The normalised polarisabilities for L 2 3 4 for HD+},
one can see that the approximation that the polarisability does not
depend on $F$ is quite good for some hyperfine states, and moderate
in others, which is due to their more or less pure character. In order
to obtain values accurate to better than one atomic unit for $\HDp$,
because of its large values of $\alpha_{t}$ it is necessary to use
the exact hyperfine dependence of the polarisability anisotropy,

The results for $\Htwop$ in odd - $L$ states are shown in Table~\ref{tab:H2+ polarisabilities}.
We can see that in this species, the anisotropic polarisabilities
are always very close to those of the pure states. The maximum deviation
is approximately 0.01 atomic unit. Thus, for current purposes, for
$\Htwop$ one may use Eq.~(\ref{eq:anisotropic polarisability any pure state})
for all rovibrational levels.

\begin{table}
$\begin{array}{|l|c|c|c|c|c|c|}
\hline \text{hyp. state} & \text{Eq.\,(\ref{eq:anisotropic polarisability any pure state})} & \text{} &  & {\rm level\,}(v,L) &  & \\
\text{\ensuremath{\ensuremath{(}}}I,\, S,\, J,\, J_{z}) & \text{(normalized)} & \text{\ensuremath{v=0}} & \text{\ensuremath{v=1}} & \text{\ensuremath{v=2}} & \text{\ensuremath{v=3}} & \text{\ensuremath{v=4}}\\
\hline \text{} & \text{} &  &  & L=1 &  & \\
\hline (1,\frac{1}{2},\frac{3}{2},0) & -1.25 & -1.24946 & -1.2495 & -1.24951 & -1.24956 & -1.2496\\
(1,\frac{1}{2},\frac{1}{2},0) & -0.125 & -0.1216 & -0.12186 & -0.12212 & -0.12238 & -0.12262\\
(1,\frac{3}{2},\frac{1}{2},0) & 2.125 & 2.12158 & 2.12185 & 2.12212 & 2.12239 & 2.12262\\
(1,\frac{3}{2},\frac{5}{2},0)* & -0.875 & -0.875 & -0.875 & -0.875 & -0.875 & -0.875\\
(1,\frac{3}{2},\frac{3}{2},0) & 1. & 0.99946 & 0.9995 & 0.99952 & 0.99956 & 0.9996\\
\hline \text{} &  &  &  & \text{\ensuremath{L=3}} &  & \\
\hline (1,\frac{1}{2},\frac{7}{2},0) & -0.75 & -0.74955 & -0.74958 & -0.74961 & -0.74964 & -0.74967\\
(1,\frac{1}{2},\frac{5}{2},0) & -0.75 & -0.74871 & -0.7488 & -0.7489 & -0.74899 & -0.74909\\
(1,\frac{3}{2},\frac{3}{2},0)* & -0.6 & -0.6 & -0.6 & -0.6 & -0.6 & -0.6\\
(1,\frac{3}{2},\frac{5}{2},0) & -0.4125 & -0.41379 & -0.41369 & -0.41359 & -0.4135 & -0.41341\\
(1,\frac{3}{2},\frac{7}{2},0) & -0.5 & -0.50044 & -0.50041 & -0.50038 & -0.50035 & -0.50032\\
(1,\frac{3}{2},\frac{9}{2},0)* & -0.6875 & -0.6875 & -0.6875 & -0.6875 & -0.6875 & -0.6875
\\\hline \end{array}$\caption{\label{tab:H2+ polarisabilities} Anisotropic polarisability of the
$J_{z}=0$ - quantum states of $\Htwop$ in $L=1,\,3$, normalized
to those of the stretched states. See caption of Tab.~\ref{tab:The-normalized-polarisability for L=00003D1 and HD+}.}
\end{table}

\clearpage{}\pagebreak{}

\section{The black-body radiation frequency shift}

\subsection{Generalities}

The black-body radiation (BBR) shift of a level $m$ is computed as

\begin{equation}
\Delta E_{BBR}(m,\, T)=-\frac{1}{2}\int_{0}^{\infty}\alpha_{s}(m,\omega){\cal E}_{BBR}(T,\omega)^{2}d\omega\ .\label{eq:BBR shift general}
\end{equation}
if the BBR electric field is unpolarized. The contributions from the
magnetic field are neglected. Therefore, under this assumption and
because of the small hyperfine splittings compared to the (smallest)
rotational levels splitting (20 MHz versus 1 THz, i.e. $2\times10^{-5}$
in relative terms), the BBR shift is to a high approximation equal
for all hyperfine states of a given rovibrational level.

\subsection{Approximate treatment}

\subsubsection{Homonuclear ions}

We may approximate the polarisability of the homonuclear ions by its
zero-frequency value: $\alpha_{s}((v,\, L),\omega)\simeq\alpha_{s}((v,\, L),\omega=0)=\alpha_{s}(v,\, L)$,
where the values are given in the Tables~\ref{table: polarisability of H2+}
and \ref{table: polarisability of D2+} above. Then

\begin{equation}
\Delta E_{BBR}(m,\, T)\simeq\Delta E_{stat}((v,\, L),\, T)=-\frac{1}{2}\alpha_{s}((v,\, L),0)\,(831.9\,{\rm V/m})^{2}\,(T/{\rm 300\, K)^{4}}.\label{eq:BBR shift - static}
\end{equation}
(In this expression, the value of $\alpha_{s}$ in atomic units is
to be multiplied by the value of $4\pi\epsilon_{0}a_{0}^{3}$ in SI
units). A polarisability of 1 atomic unit gives a frequency shift
of $-8.6\,$mHz at 300~K. The shifts of several selected rovibrational
levels are given in Table \ref{tab:Static and dynamic BBR shift of H2+}.

\begin{table}
\begin{centering}
$\begin{array}{|c|c|c|c|}
\hline v & L & \text{\ensuremath{\Delta E_{stat}((v,\, L),\, T)}} & \text{ \ensuremath{\Delta E_{dyn,elec}((v,\, L),\, T)}}\\
\text{} & \text{} & \text{[mHz]} & \text{[mHz]}\\
\hline 0 & 0 & -27.3 & -0.0023\\
0 & 1 & -27.4 & -0.0023\\
0 & 3 & -27.8 & -0.0024\\
1 & 1 & -33.7 & -0.0044\\
1 & 3 & -34.2 & -0.0046\\
2 & 1 & -41.7 & \\
3 & 1 & -51.9 & -0.0156
\\\hline \end{array}$
\par\end{centering}

\caption{\label{tab:Static and dynamic BBR shift of H2+} Static approximation
of the BBR shift and dynamic contribution of some levels of $\Htwop$,
at $T=300$~K. The total BBR shift is obtained by adding the values
in the third and fourth columns.}
\end{table}

\subsubsection{Heteronuclear ions}

\begin{table}
\caption{\label{tab:Transition dipole moments} Selected reduced transition
dipole matrix elements, $d=\left\langle v'L'\|\mathbf{d}\|vL\right\rangle $,
for transitions between rovibrational states of HD$^{+}$for the case
$L=5\to L'=6$ (in atomic units). The notation $[x]$ means $\times10^{x}$.}

\centering{}%
\begin{tabular}{|c@{\hspace{6mm}}l|c@{\hspace{6mm}}l|}
\hline
$v\to v'$ & ~~~~~~~~~~$d$ & $v\to v'$ & ~~~~~~~~~~$d$\tabularnewline
\hline
\hline
$(0\!\to\!0)$  & 0.85382285  & $(3\!\to\!0)$  & 0.22598012{[}$-$02{]} \tabularnewline
$(0\!\to\!1)$  & 0.64170498{[}$-$01{]}  & $(3\!\to\!1)$  & 0.19539417{[}$-$01{]} \tabularnewline
$(0\!\to\!2)$  & 0.97751608{[}$-$02{]}  & $(3\!\to\!2)$  & 0.20027036 \tabularnewline
$(0\!\to\!3)$  & 0.23604947{[}$-$02{]}  & $(3\!\to\!3)$  & 1.00176877 \tabularnewline
$(0\!\to\!4)$  & 0.74298137{[}$-$03{]}  & $(3\!\to\!4)$  & 0.12862769 \tabularnewline
$(0\!\to\!5)$  & 0.27963915{[}$-$03{]}  & $(3\!\to\!5)$  & 0.30336340{[}$-$01{]} \tabularnewline
\hline
$(1\!\to\!0)$  & 0.11178434  & $(4\!\to\!0)$  & 0.63859280{[}$-$03{]} \tabularnewline
$(1\!\to\!1)$  & 0.90139089  & $(4\!\to\!1)$  & 0.45000014{[}$-$02{]} \tabularnewline
$(1\!\to\!2)$  & 0.90803829{[}$-$01{]}  & $(4\!\to\!2)$  & 0.27891731{[}$-$01{]} \tabularnewline
$(1\!\to\!3)$  & 0.16798347{[}$-$01{]}  & $(4\!\to\!3)$  & 0.23541895 \tabularnewline
$(1\!\to\!4)$  & 0.46309897{[}$-$02{]}  & $(4\!\to\!4)$  & 1.05507688 \tabularnewline
$(1\!\to\!5)$  & 0.16105915{[}$-$02{]}  & $(4\!\to\!5)$  & 0.14394208 \tabularnewline
\hline
$(2\!\to\!0)$  & 0.11198057{[}$-$01{]}  & $(5\!\to\!0)$  & 0.22302649{[}$-$03{]} \tabularnewline
$(2\!\to\!1)$  & 0.16073279  & $(5\!\to\!1)$  & 0.14049130{[}$-$02{]} \tabularnewline
$(2\!\to\!2)$  & 0.95063090  & $(5\!\to\!2)$  & 0.71058396{[}$-$02{]} \tabularnewline
$(2\!\to\!3)$  & 0.11129675  & $(5\!\to\!3)$  & 0.36420347{[}$-$01{]} \tabularnewline
$(2\!\to\!4)$  & 0.23608027{[}$-$01{]}  & $(5\!\to\!4)$  & 0.26815005 \tabularnewline
$(2\!\to\!5)$  & 0.72022091{[}$-$02{]}  & $(5\!\to\!5)$  & 1.11088687 \tabularnewline
\hline
\end{tabular}
\end{table}

For the heteronuclear ions, we express the polarisability as

\begin{equation}
\alpha_{s}((v,\, L),\,\omega)=\alpha_{s}((v,\, L),\,0)+\delta\alpha_{s,dyn,rv}((v,\, L),\,\omega)+\delta\alpha_{s,dyn,elec}((v,\, L),\,\omega)\,.\label{eq:alpha_s split into parts}
\end{equation}
Here, $\delta\alpha_{s,dyn,elec}((v,\, L),\,\omega)$ is the frequency-dependent
contribution from the excited electronic levels. It does not include
the frequency-independent part, which is instead included in $\alpha_{s}((v,\, L),\,0)$.
Both $\delta\alpha_{s,dyn,elec}((v,\, L),\,\omega)$ and $\delta\alpha_{s,dyn,rv}((v,\, L),\omega)$
are defined so that they vanish at $\omega=0$. The frequency-dependent
contributions from E1 rovibrational transitions within the ground
electronic state are important and give rise to \cite{kool},
\begin{align}
\nonumber \\
\delta\alpha_{s,dyn,rv}((v,\, L),\,\omega) & =\frac{1}{3}\frac{1}{2L+1}\sum_{v',L'}|\langle v'L'||{\bf d}||v,L\rangle|^{2}\left(\frac{1}{E_{v'L'}-E_{vL}+\hbar\omega}+\frac{1}{E_{v'L'}-E_{vL}-\hbar\omega}-\frac{2}{E_{v'L'}-E_{vL}}\right)\,.\label{eq:approximation to alpha (omega)}
\end{align}
In this sum, the value of $L'$ can only take on the values $L\pm1$,
due to the selection rule. Again, $\alpha_{s}((v,L),\,0)$ is the
variational calculation result. As a first approximation, we can neglect
$\delta\alpha_{s,dyn,elec}$, as done above for the homonuclear ions,
since the transitions to the excited electronic states are of similar
character. This neglect will be corrected in the next subsection.

The total BBR shift is

\begin{equation}
\Delta E_{BBR}((v,\, L),\, T)=\Delta E_{stat}((v,\, L),\, T)+\Delta E_{dyn,rv}((v,\, L),\, T)+\Delta E_{dyn,elec}((v,\, L),\, T)\ .\label{eq:total BBR shift}
\end{equation}
We first discuss the dynamic \textit{rovibrational} contribution to
the BBR shift,

\[
\Delta E_{dyn,rv}((v,\, L),\, T)=-\frac{1}{2}\int_{0}^{\infty}\delta\alpha_{s,dyn,rv}((v,\, L),\omega){\cal E}_{BBR}(T,\omega)^{2}d\omega\ ,
\]
which we have computed for levels up to $v_{max}=10,\, L_{max}=5$,
extending the results of Ref.~\cite{kool}, which considered levels
with $v_{max}=7,\, L_{max}=1$.

In this computation, it is important to use the most accurate transition
dipoles values available, in order to reach a sufficient absolute
accuracy in the polarisability and BBR shift, since partial cancellations
occur in Eq.~(\ref{eq:approximation to alpha (omega)}). For $v<6,\, L<6$
we use the precise transition dipoles of Tian et al.~\cite{Tian et al},
based on variational wave functions. Their fractional inaccuracy is
stated as smaller than $1\times10^{-6}$, and is less than that of
our previously published values in Ref.~\cite{Zee2}. As a check,
we have recomputed the transition dipole moment of $(v=0,\, L=0)\rightarrow(v'=0,\, L'=1)$
with a larger basis set, and the value 0.3428334 at.~u. in agreement
with Tian et al. to better than $3\times10^{-7}$ in fractional terms.
We have computed the transition dipole moments between $L=5$ and
$L'=6$ levels having $v,\, v'<6$ in order to extend the results
of Tian et al. They are listed in Tab. \ref{tab:Transition dipole moments}.
For larger $v,\, v'\ge6$ we use the Born-Oppenheimer transition dipole
elements given in Ref.~\cite{hfi}. These agree, in the $v,\, L$
range computed by Tian et al., within 1 to 2 parts in $10^{4}$ with
their results. As energy differences $E_{v,L}-E_{v',L'}$ we use the
precise energies including QED corrections \cite{Korobov relativistic corrections 2006,Korobov and Zhong}
when $v<5,\, L<5$, and otherwise the values of Moss \cite{Moss 1993}.

Table \ref{tab:Values-of-the dynamic contribution to the BBR shift}~(a)
presents the relative value of the dynamic rovibrational contribution.
We see that for the $L=0$ levels a strong cancellation between the
(particularly large) contributions $\Delta E_{stat}((v,\, L=0),\, T)$
and $\Delta E_{dyn,rv}((v,\, L=0),\, T)$ occurs, which results in
a small BBR shift. In absolute terms, the BBR shift value is seen
to grow with $v$ and with $L$; see part (b) of the Table. The absolute
values are in the range of 1~mHz to several tens of mHz, for $v=0\ldots6$
and moderate $L$.

For $v<6$ we estimate the inaccuracy of $\Delta E_{dyn,rv}((v,\, L),\,300\,{\rm K})$
to be less than $10^{-7}$~Hz, since the individual contributions
to the sum are less than 0.1~Hz in absolute value. The values of
$\Delta E_{stat}((v,\, L),\,300\,{\rm K})$ are smaller than 0.1~Hz
in absolute value, and their inaccuracy is determined by the inaccuracy
of our $\alpha_{s}((v,\, L),\,\omega=0)$ values. The inaccuracy is
thus less than $10^{-6}$ Hz. However, the non-relativistic approximation
implies that both $\alpha_{s}((v,\, L),\,\omega=0)$ and the transition
dipoles are only accurate to the $1\times10^{-4}$ fractional level.
Then , the theoretical inaccuracy of the BBR shift, assuming the last
term in Eq.~(\ref{eq:total BBR shift}) is negligible, may be stated
conservatively as less than $3\times10^{-5}$~Hz for the levels $v<6$,
since the shift and its uncertainty is mostly determined by three
contributions, each with approximate uncertainty of $1\times10^{-5}$~Hz.
For $v\ge6$, taking into account that the transition dipoles values
are calculated in Born-Oppenheimer approximation, the overall inaccuracy
is estimated at $6\times10^{-5}$~Hz.

From an experimental point of view, the temperature derivative of
the BBR shift is an important quantity, since the temperature of the
BBR field in an ion trap has a relatively large uncertainty, due to
the difficulty in determining it experimentally. For $\Htwop$ this
derivative can be trivially obtained from Eq.~(\ref{eq:BBR shift - static}),
while the results for $\HDp$ are given in Tab.~\ref{tab:Values-of-the dynamic contribution to the BBR shift}~(c,
d). We find a strong variation between levels. Only for levels having
larger $v$ and $L$ the normalized derivative is close to the value
$4/T$ corresponding to a purely static BBR shift, Eq.~(\ref{eq:BBR shift - static}).

\begin{table}
\begin{centering}
(a)~~$\begin{array}{|c|cccccc|}
\hline v & \text{} &  & L &  &  & \\
\text{} & 0 & 1 & 2 & 3 & 4 & 5\\
\hline 0 & -1.0025 & -1.1333 & -0.9720 & -0.7993 & -0.6340 & -0.4865\\
1 & -1.0014 & -1.0468 & -0.9087 & -0.7593 & -0.6148 & -0.4844\\
2 & -1.0004 & -0.9606 & -0.8436 & -0.7154 & -0.5902 & -0.4761\\
3 & -0.9994 & -0.8754 & -0.7769 & -0.6682 & -0.5609 & -0.4621\\
4 & -0.9983 & -0.7917 & -0.7098 & -0.6186 & -0.5277 & -0.4432\\
5 & -0.9973 & -0.7104 & -0.6431 & -0.5674 & -0.4913 & -0.4199\\
6 & -0.9959 & -0.6322 & -0.5769 & -0.5153 & -0.4532 & -0.3925\\
7 & -0.9948 & -0.5567 & -0.5132 & -0.4636 & -0.4120 & -0.3645\\
8 & -0.9935 & -0.4864 & -0.4520 & -0.4121 & -0.3713 & -0.3330\\
9 & -0.9921 & -0.4200 & -0.3941 & -0.3635 & -0.3312 & -0.2996\\
10 & -0.9905 & -0.3596 & -0.3385 & -0.3173 & -0.2918 & -0.2672
\\\hline \end{array}$
\par\end{centering}

\begin{centering}
(b) $\begin{array}{|c|cccccc|}
\hline v & \text{} &  & L &  &  & \\
\text{} & 0 & 1 & 2 & 3 & 4 & 5\\
\hline 0 & \,\,0.0084 & \,\,0.0046 & -0.0010 & -0.0070 & -0.0129 & -0.0183\\
1 & \,\,0.0057 & \,\,0.0019 & -0.0037 & -0.0099 & -0.0160 & -0.0216\\
2 & \,\,0.0019 & -0.0019 & -0.0075 & -0.0138 & -0.0201 & -0.0261\\
3 & -0.0034 & -0.0071 & -0.0128 & -0.0192 & -0.0257 & -0.0319\\
4 & -0.0106 & -0.0143 & -0.0200 & -0.0265 & -0.0332 & -0.0397\\
5 & -0.0203 & -0.024 & -0.0297 & -0.0364 & -0.0433 & -0.0502\\
6 & -0.0353 & -0.0372 & -0.0431 & -0.0498 & -0.0570 & -0.0644\\
7 & -0.0531 & -0.0554 & -0.0612 & -0.0682 & -0.0758 & -0.0834\\
8 & -0.0780 & -0.0802 & -0.0863 & -0.0937 & -0.1017 & -0.1100\\
9 & -0.1123 & -0.1151 & -0.1213 & -0.1290 & -0.1379 & -0.1475\\
10 & -0.1609 & -0.1645 & -0.1715 & -0.1795 & -0.1897 & -0.2009
\\\hline \end{array}$
\par\end{centering}

\begin{centering}
(c)~$\begin{array}{|c|cccccc|}
\hline v &  &  & L &  &  & \\
\text{} & 0 & 1 & 2 & 3 & 4 & 5\\
\hline 0 & -0.12 & -0.13 & -0.15 & -0.17 & -0.19 & -0.22\\
1 & -0.18 & -0.18 & -0.20 & -0.22 & -0.25 & -0.28\\
2 & -0.24 & -0.25 & -0.27 & -0.29 & -0.31 & -0.35\\
3 & -0.33 & -0.34 & -0.35 & -0.38 & -0.4 & -0.44\\
4 & -0.44 & -0.45 & -0.47 & -0.49 & -0.52 & -0.55\\
5 & -0.59 & -0.60 & -0.62 & -0.64 & -0.67 & -0.71\\
6 & -0.81 & -0.79 & -0.81 & -0.84 & -0.87 & -0.92\\
7 & -1.07 & -1.05 & -1.08 & -1.11 & -1.15 & -1.19\\
8 & -1.42 & -1.41 & -1.43 & -1.47 & -1.52 & -1.57\\
9 & -1.89 & -1.89 & -1.92 & -1.97 & -2.02 & -2.10\\
10 & -2.56 & -2.58 & -2.62 & -2.67 & -2.74 & -2.84
\\\hline \end{array}\,\,\,\,\,\,{\rm (d)}\,\begin{array}{|c|cccccc|}
\hline v &  &  & L &  &  & \\
\text{} & 0 & 1 & 2 & 3 & 4 & 5\\
\hline 0 & -1.11 & -2.14 & 11.27 & 1.79 & 1.12 & 0.91\\
1 & -2.32 & -7.23 & 3.99 & 1.66 & 1.15 & 0.95\\
2 & -9.61 & 9.94 & 2.64 & 1.56 & 1.17 & 0.99\\
3 & 7.39 & 3.55 & 2.07 & 1.47 & 1.18 & 1.02\\
4 & 3.16 & 2.37 & 1.75 & 1.39 & 1.17 & 1.05\\
5 & 2.18 & 1.87 & 1.55 & 1.32 & 1.16 & 1.06\\
6 & 1.72 & 1.60 & 1.42 & 1.26 & 1.15 & 1.07\\
7 & 1.50 & 1.43 & 1.32 & 1.22 & 1.13 & 1.07\\
8 & 1.36 & 1.32 & 1.24 & 1.18 & 1.12 & 1.07\\
9 & 1.27 & 1.23 & 1.19 & 1.14 & 1.10 & 1.07\\
10 & 1.20 & 1.18 & 1.15 & 1.11 & 1.09 & 1.06
\\\hline \end{array}$
\par\end{centering}

\centering{}\caption{\label{tab:Values-of-the dynamic contribution to the BBR shift} (a)
Values of the dynamic vibrational contribution to the BBR shift of
a level $(v,\, L)$ of $\HDp$, normalized to the static contribution,
$\Delta E_{dyn,rv}((v,\, L),\, T)/\Delta E_{stat}((v,\, L),\, T)$,
at $T=300$~K. (b) Approximate total BBR shift $\Delta E_{stat}((v,\, L),\, T)+\Delta E_{dyn,rv}((v,\, L),\, T)$,
in Hz. (c) temperature derivatives of the approximate total BBR shift,
$d(\Delta E_{stat}((v,\, L),\, T)+\Delta E_{dyn,rv}((v,\, L),\, T))/dT$,
at 300~K, in mHz/K; (d) the normalized temperature derivative of
the total BBR shift, $(T/4)\times(\Delta E_{stat}+\Delta E_{dyn,rv})^{-1}\times$
$d(\Delta E_{stat}+\Delta E_{dyn,rv})/dT$, at 300~K. These results
are non-relativistic and do not include the frequency-dependent contributions
$\Delta E_{dyn,elec}$ from excited electronic states, which are given
in Tab.~\ref{tab:Contribution-to-the BBR shift of HD+ due to excited states}.
}
\end{table}

\subsection{Variational results}

\begin{figure}
\centering{}\includegraphics[scale=0.55]{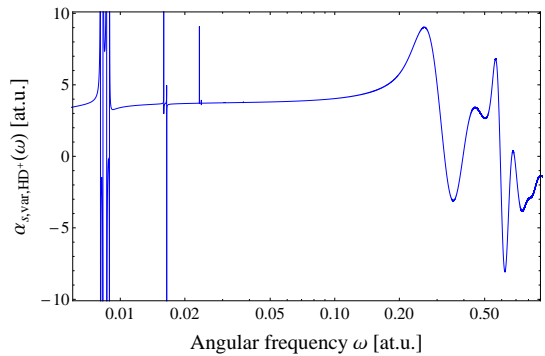}\caption{\label{fig:Plot of dynamic polarisability of HD+ in one particular level}.
The scalar polarisability of $\HDp$ in the level $(v=1,\, L=1)$,
computed using variational wave functions. Atomic units are used.}
\end{figure}

For several levels of both $\Htwop$ and $\HDp$ we have computed
the dynamic polarisability $\alpha_{s,var}(\omega)$ (and $\alpha_{t,var}(\omega)$)
\textit{directly}, using variational wave functions. For one particular
level of $\HDp$, the polarisability $\alpha_{s}$ has been computed
up to large frequencies, see Fig.~\ref{fig:Plot of dynamic polarisability of HD+ in one particular level}.
The calculation was performed using the complex coordinate rotation
method \cite{Rescigno and McKoy,Korobov - dynamic polarisability}.
This overview clearly shows the dominating contributions from the
rovibrational levels when $\omega$ is small whereas for large $\omega$
the excited electronic states yield a broad dispersive resonances.
The low-frequency tail of this resonance, as $\omega\rightarrow0$,
is responsible for giving rise to $\Delta E_{dyn,elec}$~.

For several other levels, the computation was performed up to an angular
frequency $\omega=0.1$~atomic units, in steps of $10^{-5}$ atomic
units.  The results are given in the additional material available
online \cite{Online data}. Since the computation was done in the
non-relativistic approximation, the fractional inaccuracy of the values
with respect to the exact values is approximately $1\times10^{-4}$.
This is then also the fractional inaccuracy of the BBR shifts computed
from this data.

\subsubsection{$\Htwop$}

For $\Htwop$, we can compare our values of the scalar polarisability
with the calculation by Pilon, who has communicated the values at
six different frequencies \cite{Plion Pirvate comm}. The values agree,
with deviations of at most $2\times10^{-6}$ atomic units in the range
$\omega\le0.08$.

\begin{figure}
\centering{}\includegraphics[scale=0.5]{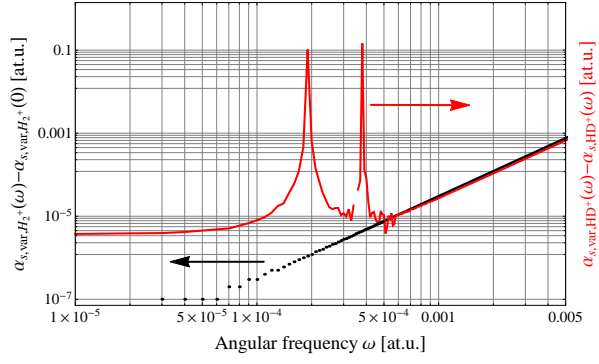}\caption{\label{fig:Plot of dynamic polarisability of HD+ and H2+}. Black
(dotted line): the frequency-dependent contribution of the scalar
polarisability of $\Htwop$ in the level $(v=1,\, L=1)$, computed
using variational wave functions. Red (gray, full line): difference
between the scalar polarisability of $\HDp$, also in the level $(v=1,\, L=1)$,
computed variationally and approximately. The difference is noticeable
close to the two rotational transition frequencies $(v=1,\, L=1)\rightarrow$$(v=1,\, L=0)$
and $(v=1,\, L=1)\rightarrow$$(v=1,\, L=2)$. A gap occurs in the
red curve at $\omega=3.5\times10^{-4}$ atomic units because the difference
is negative.}
\end{figure}

We show in Fig.~\ref{fig:Plot of dynamic polarisability of HD+ and H2+}
the frequency-dependent part of the polarisability of one level of
$\Htwop$, $\alpha_{s,var,{\rm H}_{2}^{+}}((1,\,1),\,\omega)-\alpha_{s,{\rm var,H}_{2}^{+}}((1,\,1),\,0)$,
at low frequencies. For the computation of the BBR shift at 300~K,
frequencies up to approximately $\omega=0.013$~atomic units are
relevant. In this range the polarisability is quite close to quadratic
in $\omega$. With increasing vibrational quantum numbers $v$, the
deviations from quadratic are more pronounced.

The dynamic electronic BBR shift corrections $\Delta E_{dyn,elec}$
computed from the variational data (with an integration analogous
to Eq.~(\ref{eq:BBR shift general})) are shown in Table~\ref{tab:Static and dynamic BBR shift of H2+}.
We see that the correction is small in relative terms, $1\times10^{-4}$
for $v=0$, increasing to $4\times10^{-4}$ for $v=3$. It is very
weakly dependent on $L$. Nevertheless, these results show that the
dynamic contribution should not be omitted even within the non-relativistic
approximation. When it is included, the overall inaccuracy is limited
by the non-relativistic approximation to approximately $1\times10^{-4}$
fractionally. For $v<6,\, L<6$ the total BBR shift is smaller than
0.1~Hz. Therefore, the absolute error is less than 0.01~mHz.

\subsubsection{$\HDp$}

For $\HDp$, a comparison of the variational dynamic polarisability
$\alpha_{s,var,{\rm HD}^{+}}(\omega)$ with the approximation $\alpha_{s,{\rm HD^{+}}}(\omega)=\alpha_{s}(\omega=0)+\delta\alpha_{s,dyn,rv}(\omega)$
is depicted in Fig.~\ref{fig:Plot of dynamic polarisability of HD+ and H2+},
which shows their difference. In evaluating the approximation, we
have used both the transition dipoles of Tian et al.~\cite{Tian et al}
and their non-relativistic energies, since also the variational polarisability
was computed in non-relativistic approximation. The agreement is very
good, except for small deviations near the transition frequencies
(whose nominal contribution to the BBR shift is only of order $1.5\times10^{-3}$~mHz),
and a frequency-dependent contribution from the excited electronic
states, which is again closely quadratic in frequency.

\begin{table}
\begin{centering}
$\begin{array}{|c|c|c|c|c|}
\hline v & L & \omega_{min}(v,\, L) & \text{fit} & \text{\ensuremath{\Delta E_{dyn,elec}((v,\, L),\, T)}}\\
 &  &  &  & [{\rm mHz}]\\
\hline\hline 0 & 0 & 0.0015 & 16.10\,\omega^{3}+14.54\,\omega^{2} & -0.0021\\
0 & 1 & 0.0015 & 16.24\,\omega^{3}+14.62\,\omega^{2} & -0.0021\\
0 & 3 & 0.002 & 16.94\,\omega^{3}+15.01\,\omega^{2} & -0.0022\\
0 & 4 & 0.003 & 17.58\,\omega^{3}+15.32\,\omega^{2} & -0.0022\\
1 & 1 & 0.0015 & 48.34\,\omega^{3}+24.76\,\omega^{2} & -0.0036\\
1 & 5 & 0.019 & 63.87\,\omega^{3}+25.86\,\omega^{2} & -0.0038\\
2 & 4 & 0.02 & 148.1\,\omega^{3}+42.73\,\omega^{2} & -0.0063\\
2 & 5 & 0.019 & 166.0\,\omega^{3}+42.92\,\omega^{2} & -0.0064\\
3 & 2 & 0.04 & 395.8\,\omega^{3}+62.68\,\omega^{2} & -0.0094
\\\hline \end{array}$
\par\end{centering}

\centering{}\caption{\label{tab:Contribution-to-the BBR shift of HD+ due to excited states}
Fourth column: polynomial approximation to $\alpha_{s,dyn,elec}(\omega)$,
the frequency-dependent part of the contribution to the polarisability
$\alpha_{s}$ of $\HDp$ stemming from the excited electronic states.
Fifth column: corresponding contribution to the BBR shift at 300~K.
The angular frequencies $\omega$ and $\omega_{min}$ are in atomic
units.}
\end{table}
We have fitted a simple quadratic plus cubic polynomial to the difference
$\alpha_{s,var}(\omega)-\alpha_{s,{\rm HD^{+}}}(\omega)$ between
variational and approximate frequency-dependent polarisability, over
the frequency range $\omega_{min}(v,\, L)$ to 0.05~at.~u. Here,
$\omega_{min}(v,\, L)$ is chosen appropriately so as to allow an
accurate fit. These fits represent an approximation to $\delta\alpha_{s,dyn,elec}((v,\, L),\,\omega)$
for frequencies from 0 to 0.05~at.~u. The fits are shown in Tab.~\ref{tab:Contribution-to-the BBR shift of HD+ due to excited states}.
The contribution of the cubic term is seen to be small compared to
the quadratic one for the range of frequencies relevant for the BBR
shift at 300~K. Tab.~\ref{tab:Contribution-to-the BBR shift of HD+ due to excited states}
gives the corresponding contributions to the BBR shift, to be added
to the other two contributions given in Tab.~\ref{tab:Values-of-the dynamic contribution to the BBR shift}.
The error in the values of $\Delta E_{dyn,elec}$ due to this fit
treatment is on the order of 0.001~mHz. We see that this BBR shift
contribution again varies weakly with $L$, but significantly with
$v$ and that for levels with $v=3$ it reaches $1\times10^{-5}$~Hz.
Therefore, it needs to be taken into account even within the non-relativistic
approximation, if no loss of accuracy is desired. When this is done,
the total error of the BBR shift due to the non-relativistic approximation
is expected to be $1\times10^{-4}$ fractionally, or less than $0.03\,$mHz
for the low-lying levels of $\HDp$, $v<6$.

\section{Conclusion}

We have computed the non-adiabatic static polarisabilities of the
molecular hydrogen ions $\HDp$, $\Htwop$, and $\Dtwop$, extending
significantly previous results, mostly limited to rovibrational levels
with rotational angular momentum $L=0,\,1$. For a number of rovibrational
levels, we have also computed the frequency-dependent non-adiabatic
polarisability.

The dependence of the polarisabilities on the hyperfine state has
been derived and discussed in detail. We have pointed out the special
case of the pure states, for which a simple analytical result has
been derived. This result is actually a very good approximation for
all hyperfine states of $\Htwop$. The hyperfine-state-dependence
is of crucial importance if a detailed understanding of the systematic
shifts of transition frequencies is to be performed.

We have also computed the shifts induced by the black-body radiation
field, and their temperature derivatives.

Emphasis has been given here to achieve high numerical accuracy. The
effective relative inaccuracy of our computed values is about $1\times10^{-4}$
due to the neglect of relativistic corrections. For for $\Htwop$
and $\Dtwop$ this translates in an absolute inaccuracy of 0.001 at.~u.
for all levels with $v<6,\, L<5$. For $\HDp$ in $L=0,\,1$ levels
the inaccuracy is less than 0.1 at.~u., and in $L\ge2$, it is less
than 0.003 at.~u.. An inaccuracy of 0.1 atomic unit is sufficiently
low to allow evaluating the Stark shift with a theoretical error corresponding
to the $10^{-18}$ fractional frequency level, given the typical electric
field values in ion traps.

In order to obtain accurate values of the black-body radiation shift,
we have used accurate values of the transition dipoles and we have
analyzed the importance of the contributions from excited electronic
states. We estimate the inaccuracy of the shifts to be less than 0.03~mHz
for levels with $v<6,\, L<6$, at 300~K, for both $\HDp$ and $\Htwop$.
This corresponds to theoretical fractional frequency errors on the
order of $1\times10^{-18}$.

Using the present results it becomes possible to identify theoretically
transitions having low sensitivity to external fields \cite{Bakalov and Schiller 2013,Schiller Bakalov Korobov PRL 2014}.
This represents an important aspect in the future spectroscopy of
the simplest stable molecules.\medskip{}

\thanks{We thank Dr. Olivares Pilón for communicating unpublished results.}


\begin{thebibliography}{10}
\bibitem{Online data} Online at: ({*}{*}{*}{*} to be filled by editor
{*}{*}{*}{*}). Each line of the data files has the format: $\omega$
(in atomic units), $-\alpha_{s}((v,\, L),\,\omega)$ (in atomic units),
$-\alpha_{t}((v,\, L),\,\omega)$ (in atomic units). Note that for
$\HDp$and $v=3,\, L=2$ there is a difference of 0.0001~at.~u.
between the $\omega=0$ value of the data file and the value in \ref{table: polarisability of HD+}.
This arises from the numerical procedure used, which was optimized
towards good accuracy for high $\omega$ up to the continuum threshold.
Note that the values of $\alpha_{s},\,\alpha_{t}$ in the files should
be multiplied with $(-1)$ for actual computations.
\end{thebibliography}
\end{document}